\documentclass[twocolumn,showpacs,preprintnumbers,amsmath,amssymb]{revtex4}

\usepackage{graphicx}
\usepackage{dcolumn}
\usepackage{bm}

\begin{document}

\title{Conversion of an Atomic Fermi Gas to a Long-Lived Molecular Bose Gas}

\author{Kevin E. Strecker}
\author{Guthrie B. Partridge}
\author{Randall G. Hulet}
\affiliation{Department of Physics and Astronomy and Rice Quantum
Institute, Rice University, Houston, Texas 77251}

\date{July 21, 2003}

\begin{abstract}
We have converted an ultracold Fermi gas of $^6$Li atoms into an
ultracold gas of $^6$Li$_2$ molecules by adiabatic passage through
a Feshbach resonance.  Approximately $1.5 \times 10^5$ molecules
in the least-bound, $v = 38$, vibrational level of the X$^1 \Sigma
^+_g$ singlet state are produced with an efficiency of 50\%.  The
molecules remain confined in an optical trap for times of up to 1
s before we dissociate them by a reverse adiabatic sweep.
\end{abstract}


\maketitle

Feshbach resonances have emerged as a major tool for altering the
strength and sign of interactions in ultracold atomic gases.  A
Feshbach resonance is a collisional resonance between pairs of
free atoms and a bound state of the diatomic molecule, for which
differences between the atomic and molecular magnetic moments
enable the resonance to be magnetically tuned \cite{Stwalley76a}.
Molecules can be formed near a Feshbach resonance and have
recently been detected in ultracold gases for fields close to a
resonance \cite{Donley02}. An adiabatic sweep of the magnetic
field through the Feshbach resonance has been proposed as a
highly-efficient method for converting ultracold atoms into
ultracold molecules \cite{Timmermans99, vanAbeelen99, Mies00,
Yurovsky00, Yurovsky03, Mackie02a}, as was recently demonstrated
with a Fermi gas of $^{40}$K atoms \cite{Regal03}. This method
might be used to create a Bose-Einstein condensation (BEC) of
molecules. There is heightened interest in the case where the
initial atoms are fermions \cite{Timmermans01}, since there are
close connections between Cooper pairing in the BCS theory and BEC
of molecules in the strong coupling limit \cite{Randeria95}.
However, molecules produced by this method are vibrationally
excited, and thus far, the observed molecular lifetimes have been
only $\sim$1 ms \cite{Wynar00, Regal03}. This is likely to be
shorter than the time needed for effective evaporative cooling, or
for the thermal equilibration necessary for the molecules to Bose
condense. In this paper, we report the efficient conversion of an
atomic Fermi gas of $^6$Li atoms into a gas of molecules with an
observed lifetime of $\sim$1 s.

Major components of the apparatus have been described previously
\cite{Truscott01, Strecker02}.  After accumulating approximately
10$^{10}$ bosons ($^7$Li) and 10$^9$ fermions ($^6$Li) in a
magneto-optical trap, the atoms are optically pumped into the $F =
2, m_F = 2$ and $F = 3/2, m_F = 3/2$ states, respectively, and
transferred to a magnetic trap. In contrast to our previous work,
where only the $^7$Li atoms were evaporated and the $^6$Li were
cooled sympathetically \cite{Truscott01}, we now evaporate both
isotopes. This ``dual evaporation" scheme is far more efficient,
resulting in a 100-fold increase in the number of $^6$Li atoms to
$N = 7 \times 10^7$ and a three-fold lowering of the relative
temperature to $T = 0.1~T_F$, where $T_F$ is the Fermi
temperature.  This is the largest number of trapped fermions
cooled to temperatures below $T_F$ reported thus far.  The states
of interest are not magnetically trappable, so we transfer them to
an optical trap \cite{Strecker02}. The optical trapping potential
is approximately harmonic radially, with a frequency of $\sim$800
Hz for $^6$Li and a depth of $\sim$10 $\mu$K. The potential is
box-like axially with a length of 480 $\mu$m and a depth of
$\sim$7 $\mu$K. The $^7$Li atoms are removed from the optical trap
by a resonant laser pulse.

Feshbach resonances occur between the two lowest hyperfine levels,
$F = 1/2, m_F = 1/2$ and $F = 1/2, m_F = -1/2$ in $^6$Li as was
predicted in ref. \cite{Houbiers98} and recently observed
\cite{Ohara02}. An adjustable uniform magnetic bias field of
$\sim$549 G is applied to tune near a resonance. A frequency-swept
microwave pulse transfers the $^6$Li atoms from the $F = 3/2, m_F
= 3/2$ to the $F = 1/2, m_F = 1/2$ state with nearly 100\%
efficiency.  The transition between the $F = 1/2, m_F = 1/2$ and
$F = 1/2, m_F = -1/2$ states is driven by an RF pulse at $\sim$76
MHz.  Symmetry requires that the two state mixture be incoherent,
otherwise the atoms remain indistinguishable and do not interact.
We observe a remarkable degree of coherence in this system,
including Rabi oscillations that persist for several seconds
without evidence of decoherence. In order to create an incoherent
mixture of equal populations in each state, a small magnetic
gradient is applied and the transition is driven for 1 s by a
frequency-swept triangle wave that is swept back and forth across
the resonance 50 times. Decoherence of this mixture is readily
verified by observation of rapid loss from the shallow optical
trap as the gas thermalizes. The mixture does not decohere without
the field gradient.  At the end of this process, we estimate the
number of atoms in each spin state to be $3 \times 10^5$, while
the corresponding peak density of each state is $3 \times 10^{12}$
cm$^{-3}$, giving $T_F \approx 1.4~\mu$K. Although accurate
determinations of the temperature is limited by the flat axial
density distribution in the optical trap, it is expected to be
quite low due to rapid evaporative cooling.  The radial profiles
are consistent with temperatures below $0.1 ~T_F$.

\begin{figure}
\begin{center}
\includegraphics[bb= 59 63 569 725, clip=true, angle= -90, width=1\linewidth]{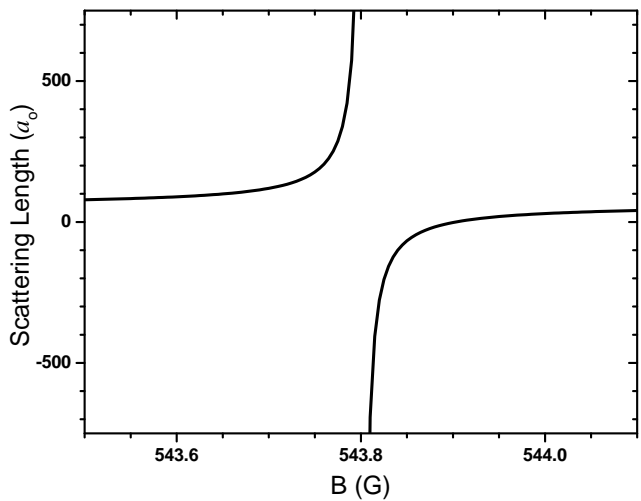}
\caption[Fig. 1]{Coupled channels calculation of the narrow
Feshbach resonance between the two lowest Zeeman sublevels of
$^6$Li.  The scattering length is shown in units of the Bohr
radius. The predicted location of the resonance is at a slightly
higher field than observed (Figs. 2 and 4).}
\end{center}
\end{figure}

Although $^6$Li exhibits an extremely large Feshbach resonance at
$\sim$850 G, we instead utilize the narrow resonance located at
$\sim$543 G.  We chose the narrow resonance for two reasons:
first, it is technically simpler to sweep over a smaller range of
magnetic field; and second, recent theoretical results indicate
that one of the primary mechanisms for loss of atoms, the creation
of pairs of hot atoms during the sweep, is minimized for a narrow
resonance compared with a broad one \cite{vanAbeelen99,
Yurovsky03}. Figure 1 shows the results of a coupled-channels
calculation of the scattering length for this resonance.  The
singlet and triplet potentials necessary for this calculation were
constructed from spectroscopic data \cite{Abraham97} and were
further refined by the observation of the location of a Feshbach
resonance in $^7$Li \cite{Strecker02}. The predicted location of
the resonance, 543.8 G, deviates slightly from the measured
location given by the data shown in the following figures. Figure
2 shows the lifetime of the trapped atoms for fixed fields near
the resonance.  The curve is asymmetric with a slower fall-off on
the low-field side of the peak at 543.25 $\pm$ 0.05 G.

\begin{figure}
\begin{center}
\includegraphics[bb= 59 63 569 725, clip=true, angle= -90, width=1\linewidth]{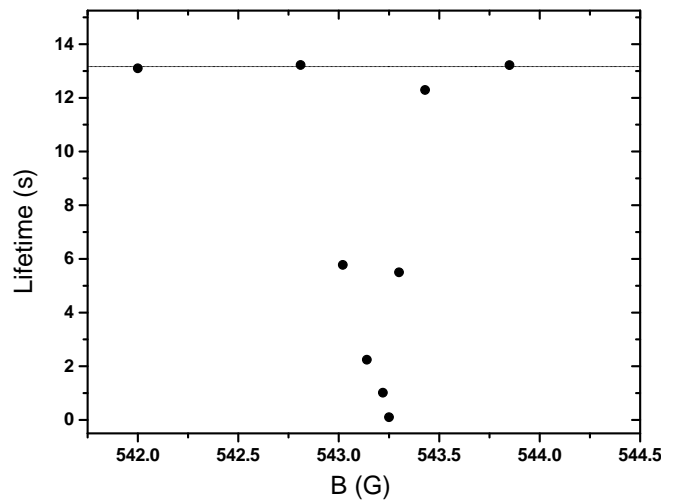}
\caption[Fig. 2]{Lifetime of an incoherent mixture of the two
lowest Zeeman sublevels as a function of the static field
strength.  The field is calibrated to 0.05 G by the frequency of
the resonance transition between the two sublevels. The lifetime
here is defined as the time at which the total atom number has
fallen to $1/e$ of its initial value.  There are no detectable
atoms at 543.25 G following the 1 s triangle wave which creates
the incoherent mixture, but by monitoring the decay of atom signal
following a short $\pi$/2 pulse instead, the lifetime is found to
be less than 200 ms. The background lifetime of 13 s, shown by the
light horizontal line, is mainly due to off-resonant scattering
induced by the trap lasers. }
\end{center}
\end{figure}

To convert atoms to molecules, the magnetic field is ramped from
high field to low so that the molecular energy goes from above
dissociation, to below. This creates molecules energetically
stable against dissociation, and minimizes the creation of
translationally hot atom pairs \cite{Timmermans99, vanAbeelen99,
Mies00, Yurovsky03, Mackie02a}. Although a Fermi gas contains a
spread of atomic energies, the process can be approximated as an
adiabatic passage through a two-level avoided crossing, where the
two states are a pair of free atoms and a bound vibrational level
of the diatomic molecule \cite{Mies00}. The Landau-Zener theory is
applicable in this case, and predicts that the transition
probability is proportional to $1 - e^{-\dot{B}^{-1}}$, where
$\dot{B}^{-1}$ is the inverse ramp rate \cite{Landau32}.
Experimentally, we start the ramp at a field far above the
resonance (549 G) and end it far below ($\sim$350 G). The measured
fraction of atoms remaining as a function of $\dot{B}^{-1}$ is
shown in Fig. 3, which exhibits the predicted exponential
dependence. The Landau-Zener model also predicts that the
conversion should be 100\% if the ramp rate is sufficiently slow,
while we observe a maximum efficiency of $\sim$50\%. A similar
maximum efficiency was reported in ref. \cite{Regal03}.

\begin{figure}
\begin{center}
\includegraphics[bb=59 63 569 725, clip=true, angle= -90, width=1\linewidth]{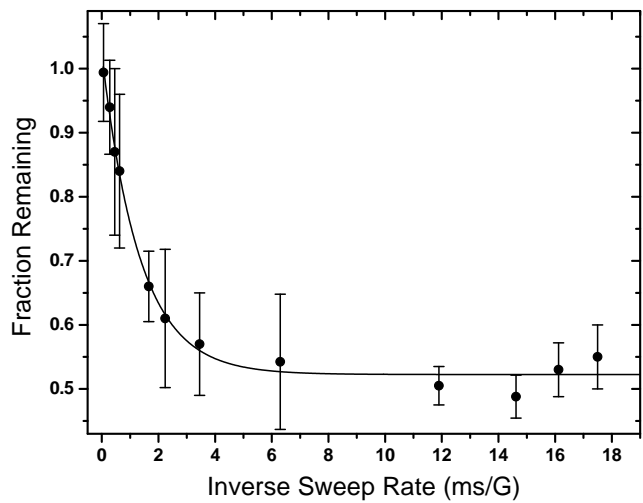}
\caption[Fig. 3]{Dependence of atom loss on inverse sweep rate.
The field is ramped linearly from high to low field. The filled
circles represent the average of 6-10 measurements and the error
bars are the standard deviations of these measurements.  The solid
line is an exponential fit, giving a decay constant of 1.3 ms/G.}
\end{center}
\end{figure}

Figure 4 gives the fraction of atoms remaining when the field ramp
begins at 549 G and is stopped at various final fields. The ramp
rate is sufficiently slow to ensure maximum conversion efficiency.
We fit the resulting data to the empirical equation $1/2 [1 + 1/(1
+ e^{-(B-B_o)/\Delta B})]$ and find that $B_o = 543.26 \, \pm \,
0.1$ G and $\Delta B = 0.23$ G.  To within our resolution of 0.1
G, which is comparable to the width of the resonance itself, the
peak of the loss (Fig. 2) and the center of the resonance, given
by $B_o$, occur at the same field.

\begin{figure}
\begin{center}
\includegraphics[bb= 59 63 569 725, clip=true, angle= -90, width=1\linewidth]{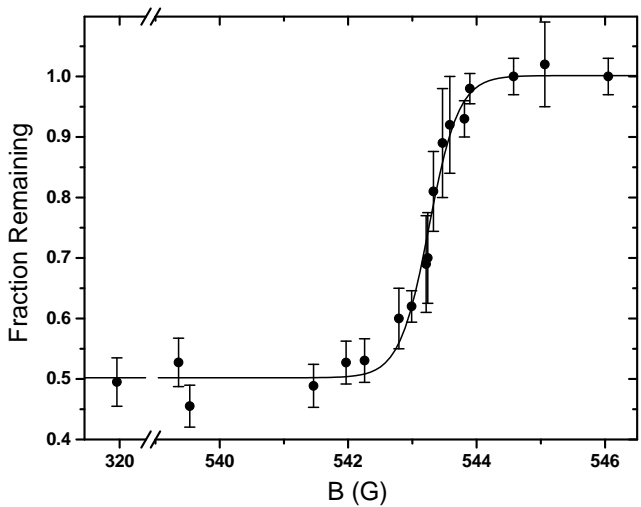}
\caption[Fig. 4]{Dependence of atom loss on final field.  The
field is ramped down from 549 G to various final fields.  The
inverse sweep rate, $\dot{B}^{-1}$, is greater than 12 ms/G for
all of the data. The data points and error bars correspond to
multiple averages, as in Fig. 3.  The solid line is a fit to an
empirical function (given in text). }
\end{center}
\end{figure}

The results presented thus far indicate that atoms are lost while
traversing the Feshbach resonance, but they do not distinguish
between molecule conversion and other loss processes, such as
inelastic collisions. To demonstrate that molecules are indeed
produced, we show that the atom/molecule conversion process is
reversible, as in ref. \cite{Regal03}, by first sweeping downward
through the resonance, waiting for a fixed amount of time, then
sweeping back to the original field. Molecules created by the
downward ramp will be returned as atoms by the upward ramp, as
long as the molecules themselves have not been lost during the
time interval $\tau$ between their creation and subsequent
dissociation back to atom pairs.  It can be seen from Fig. 5 that
close to 85\% of the original atom population survives the double
ramp, which demonstrates that no more than 15\% of the original
atom population is lost by inelastic processes. Furthermore, the
data of Fig. 5 show that the molecules survive for an unexpectedly
long time, of order 1 s.

\begin{figure}
\begin{center}
\includegraphics[bb= 59 63 569 725, clip=true, angle= -90, width=1\linewidth]{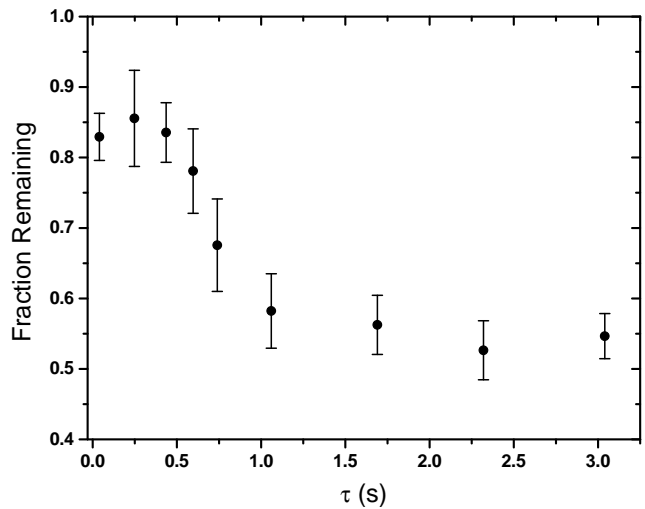}
\caption[Fig. 5]{Measurement of the molecular lifetime.  The field
is ramped downward through the Feshbach resonance and back to the
starting field.  The time $\tau$ is defined as the interval
between traversing the field $B_o$ on the downward sweep and again
on the upward sweep.  The inverse sweep rate is 3.5 ms/G and the
starting field is 549 G. The field is ramped down to a final field
whose value depends on $\tau$, and immediately ramped back to the
starting field. For the data point at 40 ms the final field is 538
G, while for $\tau$ = 248 ms, the final field is 508 G, and so on.
For the 3 points with the largest $\tau$, the field is ramped down
to 369 G and held there for the required time before being brought
back. Although we did not perform a systematic study of lifetime
as a function of field, the molecular lifetime was observed to be
field independent for the range of final fields used here.  The
data points and error bars are the averages and standard
deviations, respectively, of 8 to 16 measurements.}
\end{center}
\end{figure}

The molecules produced by this method are vibrationally excited
and can de-excite by inelastic collisions with atoms or other
molecules. These vibrational quenching collisions impart kinetic
energy to the participants in the collision equal to the energy
difference between the initial and final vibrational levels.  In
the present case, the coupled-channels calculation indicates that
the Feshbach resonance couples electronically spin-polarized pairs
of atoms interacting via the molecular triplet potential with the
least-bound ($v = 38$) vibrational level of the X$^1 \Sigma ^+_g$
singlet state of $^6$Li$_2$.  We have calculated that the $v = 37$
and $v = 38$ levels are separated by over 50 GHz, so any
vibrational quenching collision results in the release of over 2 K
of energy, far exceeding the depth of the optical trap. Previous
determinations of vibrational quenching of weakly bound
vibrational levels of the bosons $^{87}$Rb \cite{Wynar00} and Na
\cite{Stenger99, vanAbeelen99, Yurovsky00}, and of the fermion
$^{40}$K \cite{Regal03}, are consistent with lifetimes of $\sim$1
ms and rate constants of $\sim$10$^{-10}$ cm$^3 s^{-1}$.
Similarly, calculations for quenching of high-lying vibrational
levels of H$_2$ by H \cite{Balakrishnan97} and of the $v = 1$
level of Na$_2$ by Na atoms \cite{Soldan02} produce rate constants
of approximately the same value.  In contrast, we find an
effective rate constant that is two to three orders of magnitude
smaller. One might expect the quenching rate to be minimized by
stopping the field ramp within the Feshbach resonance.  In this
case, the molecules are extremely weakly bound, and will have poor
wavefunction overlap with lower-lying levels.  However, this is
not the origin of the rate reduction seen here, as the final field
is well below the resonance, and the molecules are essentially
pure singlet in character. Fermi statistics might cause a
reduction in the rate of quenching if the size of the molecule is
smaller than the size of the region over which the pair
correlation is depleted.

The long molecular lifetime may enable the formation of a
molecular BEC by allowing enough time for thermalization and
evaporative cooling of the gas.  Additionally, if the initial
Fermi gas is already at $T=0$ and the conversion is slow with
respect to the translational degrees of freedom, the conversion
should proceed along the ground state of the system directly into
a molecular BEC.  For faster sweeps that are non-adiabatic with
respect to the translational motion, but still adiabatic with
respect to molecule formation, the molecular gas will have the
same mean energy as the initial Fermi gas.  In this case, the
temperature is essentially equal to the BEC transition temperature
of the molecules.  If a molecular BEC can be formed, a completely
adiabatic sweep back across the Feshbach resonance will maintain
$T=0$, and allow the formation and investigation of Cooper pairs
at any adjustable interaction strength \cite{Timmermans01}.

The authors gratefully acknowledge the contributions of Ying-Cheng
Chen, and helpful conversations with Michael Jack and Andrew
Truscott. This work was funded by grants from the NSF, ONR, NASA,
and the Welch Foundation.

\bibliographystyle{apsrev}

\end{document}